\documentclass[twocolumn,prl,showpacs]{revtex4}%
\usepackage{amsmath}
\usepackage{amsfonts}
\usepackage{amssymb}
\usepackage{graphicx}%
\providecommand{\U}[1]{\protect\rule{.1in}{.1in}}
\providecommand{\U}[1]{\protect\rule{.1in}{.1in}}
\begin{document}
\author{D. Nissenbaum$^{1}$, L. Spanu$^{2,3}$, C. Attaccalite$^{4}$, B.
Barbiellini$^{1}$, A. Bansil$^{1}$}
\affiliation{$^{1}$Physics Department, Northeastern University, Boston MA 02115 }
\affiliation{$^{2}$INFM-Democritos, National Simulation Center and International School for
Advanced Studies (SISSA), I-34014 Trieste, Italy }
\affiliation{$^{3}$Department of Chemistry, University of California at Davis, One Shield
Avenue Davis, CA 95616}
\affiliation{$^{4}$ Universidad del Pais Vasco, Unidad de Fisica de Materiales, San
Sebastian, Spain }
\title{The Resonating-Valence-Bond Ground State of Li Nanoclusters}
\date{\today}


\begin{abstract}
We have performed Diffusion Quantum Monte Carlo simulations of Li clusters
showing that Resonating-Valence-Bond (RVB) pairing correlations between
electrons provide a substantial contribution to the cohesive energy. The RVB
effects are identified in terms of electron transfers from s- to p-like
character, constituting a possible explanation for the breakdown of the Fermi
liquid picture observed in recent high resolution Compton scattering
experiments for bulk Li.

\end{abstract}
\maketitle

Lithium, and other alkali metals, have been modeled as free-electron like
systems - an electron gas permeated by ions \cite{textbook}. Nevertheless, in
recent years, experimental and theoretical investigations of Li have revealed
a more complex phase diagram \cite{ashcroft,hanfland}. Even at ambient
pressure, the electron momentum density cannot be described adequately in
terms of Fermi liquid theory, because pronounced deviations have been observed
in bulk Li in recent high resolution Compton scattering experiments
\cite{sakurai,shulke}. Such deviations from the standard metallic picture can
be ascribed to the possible existence of significant pairing correlations in
the ground state \cite{momentum}. Bonding properties have also revealed that
Li behaves like a "bad" free-electron metal \cite{rousseau2}.

The notion of stabilizing the metallic state through the creation of a
resonant valence bond (RVB) state involving the metallic orbitals dates back
to the early works of Pauling, who first applied this picture to the Li ground
state \cite{paulingLi4}. In 1987, Anderson proposed the RVB wave function as
the natural ground state for the high temperature superconducting materials
\cite{anderson}, arguing that this picture is capable of capturing many
aspects of the phase diagram of the cuprates \cite{lee,spanu}. More recently,
it has been shown \cite{PT0408} that Pauling's RVB idea cannot account for all
of the properties of metals that depend on the existence of a Fermi Surface
(FS). However, since the nature of the FS in bulk Li has been questioned by
the high resolution Compton scattering studies \cite{shulke}, it is natural to
ask if the RVB paradigm might provide a viable model of the Li ground state.

This work demonstrates that RVB pairing correlations between electrons provide
a contribution to the total energy of Li clusters of about 20 meV or greater
per atom. The pairing correlation effects modify the electron momentum density
distribution \cite{momentum} and therefore provide a possible mechanism for
the breakdown of the Fermi liquid picture in bulk Li.

A number of authors have performed calculations on bulk Li as well as Li
nanoclusters \cite{rousseau2}, but an implementation of the RVB model for Li
clusters utilizing QMC simulations has not been attempted. A previous RVB
study of small Li clusters \cite{mcweenyLi} did not possess the accuracy of
modern QMC methodologies. Here, we report QMC calculations of correlation
effects beyond the limitations of the standard Jastrow-Slater wave function
(WF) for the Fermi liquid ground state \cite{fulde}\ by employing the RVB.
Specifically, we have used the Jastrow Antisymmetrized Geminal Product
(Jastrow+AGP, or JAGP), developed by Sorella and coworkers
\cite{casula-atoms,casula-molecules,benzene,hydrogen,acqua}, as a QMC
variational ansatz for the RVB. We have performed Diffusion Monte Carlo (DMC)
calculations to obtain precise estimates of the energy, this technique being
limited in accuracy only by the nodal structure of the variational ansatz
\cite{umrigar2007,mitas}. For a useful DMC study of bulk Li, see
\cite{filippi}. For all our Li clusters, we obtain a distinct nodal structure
improvement of the cohesive energy in comparison to standard JS WFs. An
eigenvalue analysis of the pairing wave function further confirms the RVB
nature of the ground state.%

\begin{figure}
[ptb]
\begin{center}
\includegraphics[
height=2.5633in,
width=3.2897in
]%
{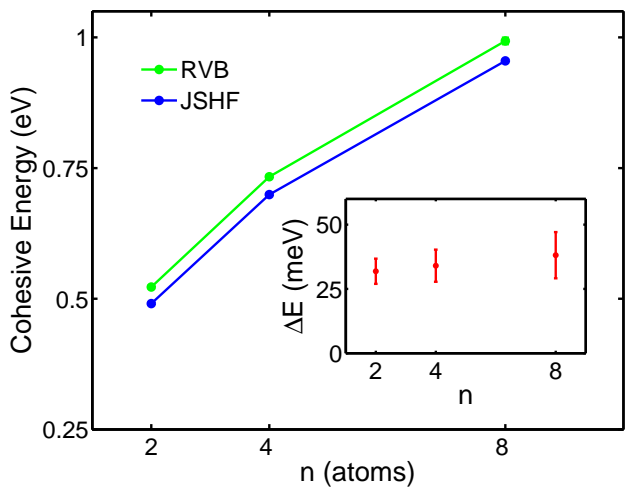}%
\caption{(Color online) Cohesive energies per atom for Li$_{n}$ with n = 2, 4
and 8, representing systems of dimension d = 1, 2 and 3, respectively,
calculated using the diffusion Quantum Monte Carlo all-electron method. The
plot compares the RVB nodal structure against the standard HF nodal structure.
The error bars are comparable to the size of the points and lines connect
points to guide the eye. The inset shows the difference in cohesive energies
between the models.}%
\label{cohesive}%
\end{center}
\end{figure}

Our calculations employ the JAGP WF, defined as the product of a Jastrow term
$J$ and an antisymmetrized determinant $\Psi_{AGP}$: $\Psi_{JAGP}%
(r_{1},...,r_{N})=\Psi_{AGP}(r_{1},...,r_{N})J(r_{1},...,r_{N})$. The
determinant $\Psi_{AGP}$ is constructed as $\Psi_{AGP}=\hat{A}[\Phi
(r_{1}^{\uparrow},r_{1}^{\downarrow})\cdots\Phi(r_{N/2}^{\uparrow}%
,r_{N/2}^{\downarrow})]$, where $\hat{A}$ is the antisymmetrization operator,
$\Phi(r_{1}^{\uparrow},r_{1}^{\downarrow})$ is the pairing function, and $N$
is the number of electrons in the system ($N$ must be even for this
formulation; see \cite{geminal} for an extension to open-shell systems). The
pairing function $\Phi$ is defined as $\Phi(r^{\uparrow},r^{\downarrow}%
)=\psi(r^{\uparrow},r^{\downarrow})1/\sqrt{2}(|\uparrow\downarrow
\rangle-|\downarrow\uparrow\rangle)$, where the spin part is a singlet. The
spatial part $\psi$ - the geminal - is represented by a pairing expansion over
a local single-particle basis set $\{\phi_{i}\}$, i.e.
\begin{equation}
\psi(r^{\uparrow},r^{\downarrow})=\sum_{i,j}\lambda_{i,j}\phi_{i}%
({r}^{\uparrow})\phi_{j}({r^{\downarrow}})\text{.}\label{geminal}%
\end{equation}
Here, indices $i$ and $j$ run over different orbitals (covering all nuclear
sites), which are expanded in a gaussian basis set centered on the nuclear
positions \cite{basis}. The Jastrow factor $J=J_{1}J_{2}J_{3}$ is composed of
an electron-nuclear ($J_{1}$), an electron-electron ($J_{2}$), and an
electron-electron-nuclear ($J_{3}$) term; it guarantees that the cusp
conditions are satisfied and it allows the correct charge distribution in the
system. The $J_{3}$ term is constructed in a form similar to the pairing
function of Eq. \ref{geminal} \cite{casula-molecules}. The Jastrow parameters,
$\lambda$ parameters, gaussian (Slater) orbital exponents, and orbital
coefficients have been optimized by energy minimization using the methods
described in Refs. \cite{sorella2005,umrigar2007}.

We have compared the JAGP wave function with three types of single-determinant
JS wave functions: one involving a standard Slater determinant of Hartree-Fock
(HF) orbitals (abbreviated JS-HF); one utilizing a pseudopotential to replace
the core electrons and utilizing Kohn-Sham orbitals for the valence electrons
calculated within the Density Functional Theory (DFT) using the Local Density
Approximation (JS-LDA); and a wave function defined as the limiting case of
the JAGP wave function in which the occupation of the virtual orbitals is
forced to be zero (JS-V0). The latter wave function possesses the standard JS
form. The geometries of all our Li clusters were optimized using the software
Jaguar with a 63111G basis set and the B3LYP exchange-correlation potential
\cite{jaguar,nissenbaum}.

For the all-electron JS-HF wave function, we examined clusters of 2, 4, and 8
Li atoms modeled with HF orbitals obtained from the software Jaguar
\cite{jaguar}. The wave function included the $J_{1}$ and $J_{2}$ terms, which
were optimized using an improved version of the Stochastic Gradient
Optimization (SGA) \cite{sga} method. Our corresponding DMC results are given
in Table ~\ref{table1} for the dimer, the planar cluster Li$_{4}$ and the
three- dimensional Li$_{8}$. For all three clusters, we obtain distinct
corrections of about 30 meV/atom for the cohesive energy. Fig.~\ref{cohesive}
illustrates the results. The DMC values with the JS wave functions are already
quite good, because the cohesive energy, $E_{coh}$, compares fairly well with
the experimental values given in Ref.~\cite{jones-Li}, and because JS DMC
calculations are known to retrieve better than 90\% of the correlation energy
\cite{foulkes}. Also, in both cases we observe the expected increase in
cohesive energy with cluster size, describing the tendency for these clusters
to grow \cite{jones-Li}. However, the inset of Fig.~\ref{cohesive} shows that
the difference $\Delta E=E_{coh}^{JAGP}-E_{coh}^{JS-HF}$, i.e. the nodal
structure corrections, are about 30 meV/atom.

\begin{table}[t]%
\begin{tabular}
[c]{|c|c|c|c|c|c|}\hline
N & $E_{DMC}^{JS-HF}$ & $E_{DMC}^{JAGP}$ & $E_{coh}^{JS-HF}$ & $E_{coh}%
^{JAGP}$ & $\Delta E_{coh}$\\
& (Hartree) & (Hartree) & (eV) & (eV) & (meV)\\\hline
2 & $-7.49593(8)$ & $-7.4971(1)$ & 0.491(2) & 0.522(3) & 32(5)\\
4 & $-7.5036(1)$ & $-7.50485(13)$ & 0.699(3) & 0.733(4) & 34(6)\\
8 & $-7.5130(1)$ & $-7.51440(23)$ & 0.955(3) & 0.993(6) & 38(9)\\\hline
\end{tabular}
\caption{DMC all-electron calculations comparing the JS-HF and the JAGP wave
function for Li clusters. The first two columns present the total energy/atom.
The next two columns show the cohesive energy/atom, and the final column is
the difference between these two cohesive energies.}%
\label{table1}%
\end{table}

\begin{table}[t]%
\begin{tabular}
[c]{|c|c|c|c|c|c|}\hline
N & $E_{DMC}^{JS-LDA}$ & $E_{DMC}^{JAGP}$ & $E_{coh}^{JS-LDA}$ &
$E_{coh}^{JAGP}$ & $\Delta E_{coh}$\\
& (Hartree) & (Hartree) & (eV) & (eV) & (meV)\\\hline
2 & $-0.21318(4)$ & $-0.21534(8)$ & 0.45932(7) & 0.51625(9) & 56.9(2)\\
4 & $-0.22100(3)$ & $-0.22357(5)$ & 0.67212(6) & 0.74020(6) & 68.1(1)\\
8 & $-0.23152(3)$ & $-0.23225(1)$ & 0.95839(6) & 0.97640(2) & 18.01(8)\\
20 & $-0.237750(5)$ & $-0.23893(1)$ & 1.12791(4) & 1.15817(2) &
30.26(6)\\\hline
\end{tabular}
\caption{Same as Table~\ref{table1}, except here DMC pseudopotential
calculations are used to compare the JS-LDA and the JAGP wave function. This
data set includes Li$_{20}$.}%
\label{table2}%
\end{table}

\begin{table}[t]%
\begin{tabular}
[c]{|c|c|c|c|c|}\hline
N & $E_{VMC}^{JS-LDA}$ & $E_{VMC}^{JAGP}$ & $\sigma_{JS-LDA}^{2}$ &
$\sigma_{JAGP}^{2}$\\
& (Hartree) & (Hartree) & (eV) & (eV)\\\hline
2 & -0.20561(4) & -0.21509(3) & 0.2921(9) & 0.068(5)\\
4 & -0.20899(4) & -0.22259(2) & 0.392(3) & 0.0477(9)\\
8 & -0.21616(2) & -0.231200(6) & 0.388(3) & 0.0662(5)\\
20 & -0.21834(2) & -0.2371(1) & 0.602(2) & 0.011(1)\\\hline
\end{tabular}
\caption{VMC calculations of the JS-LDA and the JAGP wave function for Li
clusters. The first two columns compare the total energy/atom. The variational
JAGP results for Li$_{2}$ and Li$_{4}$ are lower than the diffusion JS-LDA
results from Table~\ref{table2}, indicating superior performance of the wave
function even at the variational level for the JAGP than can be obtained at
the diffusion level for the JS-LDA. The final two columns show the variance
$\sigma^{2}$ of the local energy/atom. The JAGP shows a smaller variance than
the JS-LDA.}%
\label{table3}%
\end{table}

The bonding properties of Li can be described quite accurately by replacing
the core electrons with a pseudopotential, because the $1s$ core states do not
contribute significantly to bonding. For example, pseudopotential DMC
calculations successfully predict the small binding energy for the LiPs
molecule \cite{lips_pseudo}, in accordance with the results of more
sophisticated DMC all-electron calculations \cite{lips_all}. Another advantage
of the pseudopotential is the possibility of relatively straightforwardly
disentangling the valence characteristics of the $\Lambda=\lambda_{ij}$ matrix
in~Eq. \ref{geminal} from core contributions. This submatrix will also be seen
below to help identify an RVB signature in the wave function.

We consider the JS-LDA pseudopotential wave function for Li clusters
containing 2, 4, 8, and 20 atoms, in order to assess the impact of these
one-body orbitals on the nodal structure. The inner $1s$ core electrons were
replaced by the norm-conserving pseudopotential provided by Burkatzki \emph{et
al.} \cite{burkatzki}. The wave function was constructed using Kohn-Sham LDA
orbitals obtained with the PWscf code \cite{PWSCF}. The LDA calculations used
a cubic simulation cell of $40$ a.u. sides, a plane-wave cut-off of $70$ Ry,
and included optimized $J_{1}$ and $J_{2}$ terms. A modified version of the
DMC, the Lattice Regularalized Diffusion Monte Carlo (LRDMC) \cite{lrdmc}
method, was used, which allows the inclusion of a non-local pseudopotential in
a consistent variational scheme. The JS-LDA DMC results are summarized in
Table~\ref{table2}. In this case also, the JAGP yields the lowest total
energies. These results confirm that in the case of Li clusters the RVB nodal
structure provides a correction to the cohesive energy, which tends to be more
than 20 meV/atom.

The signature of an RVB state can be directly identified by analyzing the
eigenvalues and eigenvectors of the matrix $\Lambda=\lambda_{ij}$ in ~Eq.
\ref{geminal}. Given the $\Lambda$ matrix, we solve the eigenvalue problem
$\Lambda S\vec{u}=\mu\vec{u}$, where $S$ is the overlap matrix between the
local basis set orbitals, i.e. $S_{ij}=\langle\phi_{i}\phi_{j}\rangle$, and
$\vec{u}$ are the eigenvectors representing the basis of natural orbitals,
with corresponding eigenvalues providing the geminal coefficients associated
with the occupation numbers \cite{nso_bba}. Anderson has shown \cite{anderson}
that a signature of the RVB state is given by a change of sign of the
eigenvalues $\mu$ when passing from occupied to virtual states with a
significant weight for the virtual states. This trend also helps to reduce
double occupancy at lattice sites, as in the Heitler London limit.

\begin{figure}
[ptb]
\begin{center}
\includegraphics[
height=2.3696in,
width=3.2897in
]%
{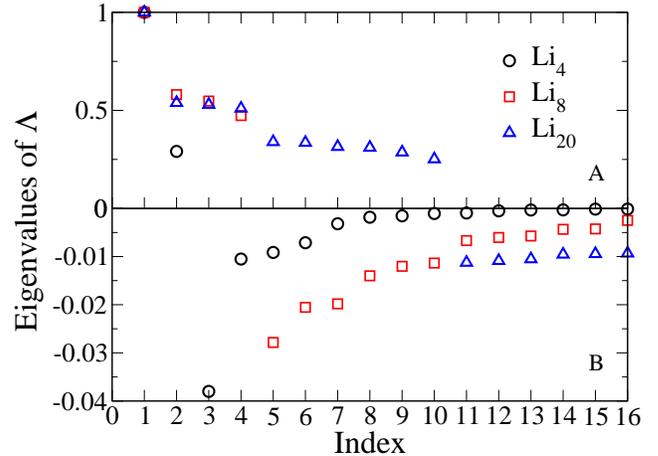}%
\caption{(Color online) Eigenvalues of the diagonalized $\Lambda$ matrix,
representing the occupation of the natural geminal orbitals. Top half:
eigenvalues for the first $\frac{N}{2}$ natural orbitals. Bottom half:
expanded view of the eigenvalues for the virtual orbitals. Unlike a standard
Jastrow-Slater wave function, the JAGP allows occupancy of the higher-level
orbitals. A characteristic of the RVB is a sign flip when passing from the
primary to the virtual orbitals.}%
\label{orbs}%
\end{center}
\end{figure}

This diagonalization was performed for Li$_{4}$, Li$_{8}$ and Li$_{20}$ with
pseudopotential. Fig. \ref{orbs} shows the eigenvalues. Positive eigenvalues
are in the top portion of the figure, while the negative eigenvalues are shown
in the bottom on an expanded scale. The presence of the RVB signature
indicates a departure from the standard JS nodal structure, and by extension
into the bulk, a departure from the Fermi liquid picture \cite{fulde}. This
departure can be explained as follows. If the generating geminal yields a
standard JS ansatz, the occupied orbitals would have equal weights, and the
remaining orbitals would not contribute. However, in the case of the RVB, the
virtual orbitals have significant non-zero amplitudes, but with opposite sign.
Contrary to the standard JS description, virtual orbitals, i.e. orbitals with
index greater than $N/2$, have a small but finite, negative eigenvalue, which
is related to the occupancy.

The behavior of the eigenvalues in Fig. \ref{orbs} confirms the fact that the
AGP structure of the Li dimer studied by Elander \emph{et al.} for various
atomic separations \cite{elander} captures general features which suggest a
certain robustness of the RVB signature with respect to the exact geometry of
Li clusters.

It is interesting to note that the natural orbitals corresponding to the
leading eigenvalues contain a significant $p$-character. An important $2p-2p$
$\pi$-contribution was observed in the AGP wave function for Li$_{2}$
\cite{elander}, causing a departure from the standard JS nodal structure.
Recent experiments find that Li impurities in an Al matrix produce an
anomalous transfer from $s$- to $p$-like character and thus constitute another
example in which the standard Fermi liquid picture breaks down and properties
of the correlated inhomogeneous electron gas must be considered \cite{alli}.%

Table \ref{table3} provides the results of Variational Monte Carlo (VMC)
calculations, rather than DMC, of the JS-LDA and the JAGP wave functions,
confirming the relatively high quality of the JAGP wave function at the
variational level. A comparison of tables \ref{table3} and \ref{table2}
indicates that our variational results using the JAGP are superior to the
diffusion results using the JS-LDA for Li$_{2}$ and Li$_{4}$.
Table~\ref{table3} also shows the significantly smaller variance of the local
energy for the RVB wave function, indicating more efficient and less
time-consuming calculations \cite{nissenbaum}, as well as a wave function that
may be approaching the exact solution. Interestingly, for the Li$_{2}$
molecule the JAGP is able to recover 95.7\% and 99.45\% of the correlation
energy \cite{casula-molecules} at the VMC and DMC levels, respectively, while
the corresponding values with backflow corrections \cite{rios} are 87.79\% and
97.1\%. Therefore, in the present case, the RVB correlations appear to
dominate other effects. Moreover, the possibility that the RVB Pauling
structures provide an important contribution in the configuration interaction
(CI) expansion has also been suggested in a recent quantum chemical study for
the neutral Li$_{4}$ cluster \cite{quintao}.

Finally, to directly measure nodal structure effects, we have utilized the
limiting JS-V0 wave function. In this way, we can estimate the energy
contribution from occupation of the virtual orbitals by forcing the pairing
wave function to possess only the $N/2$ fully occupied orbitals, with zero
occupancy of the virtual orbitals. These calculations were performed for
Li$_{4}$, Li$_{8}$ and Li$_{20}$ with pseudopotential, and in all three cases,
VMC as well as DMC calculations find an improvement of at least 20 meV/atom in
the cohesive energy due to the RVB nodal structure, consistent with the DMC
results of Table \ref{table2}. As stated previously, the definition of JS-V0
assures that the Jastrow factor, basis set, and functional form are identical
in the JAGP and in the JS-V0 wave functions, so that $\Delta E_{RVB}%
=E_{coh}^{JAGP}-E_{coh}^{JS-V0}$ is a direct calculation of the resonance
energy. This confirms that the gain in energy is a consequence of the RVB
orbitals, and not the result of a different Jastrow factor or a different
single particle basis set.

In conclusion, we have performed DMC calculations of Li clusters with an RVB
guiding WF, utilizing a fundamentally different nodal structure than the
standard JS WF. We find that the RVB nodal structure is able to recover about
20 meV of cohesive energy/atom. Furthermore, we have identified a distinct RVB
signature in an eigenvalue decomposition of the JAGP $\Lambda$-matrix,
suggesting modifications of the electron occupation numbers due to RVB
effects. These results justify use of an approximation for AGP correlation
effects in momentum density calculations \cite{momentum} and might explain
deviations from the Fermi liquid picture observed in recent high resolution
Compton scattering experiments on bulk Li.

We acknowledge useful discussions with R. S. Markiewicz. This work was
supported by the Division of Materials Science and Engineering, Basic Energy
Sciences, Office of Science of the U.S. Department of Energy contract
DE-FG02-07ER46352, and benefited from the allocation of supercomputer time at
the NERSC and the Northeastern University's Advanced Scientific Computation
Center (NU-ASCC). The work was also supported by the U.S. Department of
Energy, Scidac, contract number DE-FC02-06ER25794.


\begin{thebibliography}{99}                                                                                               %
\bibitem {textbook}N. W. Ashcroft and N. D. Mermin, \textit{Solid State
Physics}, Saunders College, Philadelphia (1976).

\bibitem {ashcroft}J. B. Nelson and N. W. Ashcroft, Nature \textbf{400} 141 (1999).

\bibitem {hanfland}M. Hanfland \textit{et. al.,} Nature \textbf{408} 174 (2000).

\bibitem {sakurai}Y. Sakurai \textit{et. al.,} Phys. Rev. Lett.\textbf{74}
2252 (1995).

\bibitem {shulke}W. Sch\"{u}lke, G. Stutz, F. Wohlert and A. Kaprolat, Phys.
Rev. B \textbf{54} 14381 (1996).

\bibitem {momentum}B. Barbiellini and A. Bansil, J. Phys. Chem. Solids
\textbf{62} 2181 (2001).

\bibitem {rousseau2}R. Rousseau and D. Marx, Chem. Eur. J. \textbf{6} 2982 (2000).

\bibitem {paulingLi4}L. Pauling, Nature \textbf{61} 1019 (1948).

\bibitem {anderson}P. W. Anderson, Science \textbf{235} 1196 (1987).

\bibitem {lee}P. A. Lee \textit{et. al.,} Rev. Mod. Phys. \textbf{78} 17 (2006).

\bibitem {spanu}L. Spanu, M. Lugas, F. Becca and S. Sorella, Phys. Rev. B
\textbf{77}, 024510 (2008).

\bibitem {PT0408}P. W. Anderson, Physics Today, April, 2008, p. 8.

\bibitem {mcweenyLi}J. R. Mohallem \textit{et. al.,} Z. Phys. D \textbf{42}
135 (1997).

\bibitem {fulde}P. Fulde, \textit{Electron Correlations in Molecules and
Solids}, Springer, Berlin (1995).

\bibitem {casula-atoms}M. Casula and S. Sorella, J. Chem. Phys. \textbf{119}
6500 (2003).

\bibitem {casula-molecules}M. Casula \textit{et. al.,} J. Chem. Phys.
\textbf{121} 7110 (2004).

\bibitem {benzene}S. Sorella \textit{et. al.,} J. Chem. Phys. \textbf{127}
014105 (2007).

\bibitem {hydrogen}C. Attaccalite and S. Sorella, Phys. Rev. Lett.
\textbf{100} 114501 (2008).

\bibitem {acqua}F.Sterpone \textit{et. al.,} J. Chem. Theory Comput.
\textbf{4} 1428-1434 (2008).

\bibitem {umrigar2007}C. J. Umrigar, J. Toulouse, C. Filippi, S. Sorella and
R. G. Hennig, Phys. Rev. Lett. \textbf{98} 110201 (2007).

\bibitem {mitas}L. Mitas, Phys. Rev. Lett. \textbf{96} 240402 (2006).

\bibitem {filippi}C. Filippi and D. M. Ceperley, Phys. Rev. B \textbf{59} 7907 (1999).

\bibitem {geminal}A. J. Coleman, J. Math. Phys. \textbf{13} 214 (1972).

\bibitem {basis}{For the all-electron case we use a Gaussian basis set of 8s6p
contracted to [3s1p], while in the psedopotential calculations a 4s4p
contracts to [2s1p].}

\bibitem {sorella2005}S. Sorella, Phys. Rev. B \textbf{71} 241103(R) (2005).

\bibitem {jaguar}http://www.schrodinger.com

\bibitem {nissenbaum}D. Nissenbaum, B. Barbiellini and A. Bansil, Phys. Rev. B
\textbf{76} 033412 (2007).

\bibitem {sga}A. Harju, B. Barbiellini, S. Siljamaki, R. M. Nieminen and G.
Ortiz, Phys. Rev. Lett. \textbf{79} 1173 (1997).

\bibitem {jones-Li}R. O. Jones \textit{et. al.,} J. Chem. Phys. \textbf{106}
4566 (1997).

\bibitem {foulkes}W. M. C. Foulkes \textit{et. al.,} Rev. Mod. Phys.
\textbf{73} 33 (2001).

\bibitem {lips_pseudo}A. Harju \textit{et. al.}, J. Radioanal. Nucl. Chem.
\textbf{21} 1931 (1996).

\bibitem {lips_all}D. Bressanini \textit{et. al.}, J. Chem. Phys. \textbf{108}
4756 (1998).

\bibitem {burkatzki}M. Burkatzki \textit{et. al.,} J. Chem. Phys. \textbf{126}
234105 (2007).

\bibitem {PWSCF}http://www.pwscf.org

\bibitem {lrdmc}M. Casula, C. Filippi and S. Sorella, Phys. Rev. Lett.
\textbf{95} 100201 (2005).

\bibitem {nso_bba}B. Barbiellini, J. Phys. Chem. Solids \textbf{61} 341 (2000).

\bibitem {elander}N. Elander, \textit{et. al.}, Intl. J. Quan. Chem.
\textbf{23} 1047 (1983).

\bibitem {alli}J. Kwiatkowska \textit{et. al.,} Phys. Rev. Lett. \textbf{96}
186403 (2006).

\bibitem {rios}LopezRios, A. Ma, N. D. Drummond, M. D. Towler and R. J. Needs,
{Phys. Rev. E} \textbf{74} 066701 (2006).

\bibitem {quintao}Quintao and R.O. Vianna, \textit{Int. J Quantum Chem.} 81,
76 (2001)
\end{thebibliography}
\end{document}